\newcommand\araa{{ARA\&A\,}}%
\newcommand\apj{{ApJ\,}}%
\newcommand\apjs{{ApJS\,}}%
\newcommand\aap{{A\&A\,}}%
\documentclass[graybox, natbib, footinfo]{svmult}

\usepackage{mathptmx}       % selects Times Roman as basic font
\usepackage{helvet}         % selects Helvetica as sans-serif font
\usepackage{courier}        % selects Courier as typewriter font
\usepackage{type1cm}        % activate if the above 3 fonts are
\usepackage{makeidx}         % allows index generation
\usepackage{graphicx}        % standard LaTeX graphics tool
\usepackage{multicol}        % used for the two-column index
\usepackage[bottom]{footmisc}% places footnotes at page bottom

\makeindex             % used for the subject index
\begin{document}
\title*{Ages of stars: Methods and uncertainties}
\author{David R. Soderblom}
\institute{David R. Soderblom \at Space Telescope Science Institute, Baltimore MD USA, \email{drs@stsci.edu}}
%\and Name of Second Author \at Name, Address of Institute \email{name@email.address}}
\maketitle

\abstract{
Estimating ages for stars is difficult at best, but Galactic problems have their own requirements that go beyond those for other areas of astrophysics.  As in other areas, asteroseismology is helping, and in this review I discuss some of the general problems encountered and some specific to large-scale studies of the Milky Way.
}

\section{Why care about ages?}
\label{sec:1}
Much of what astronomers do in studying stars and our Galaxy explicitly or implicitly involves evolutionary changes: how things change with time, and the sequence of events.  Yet time is not really a direct agent of change in stars, it is more a medium in which gradual changes occur.  Because time is not a direct agent it leaves no direct indicators and we are left estimating age in mostly indirect -- and so inexact -- ways.

The problem is illustrated by considering the Sun.  It is the only star for which we have a fundamental age, one for which all the physics is fully understood and all needed measurements can be made, and its age of $4,567\pm1\pm5$ Myr is both precise and exact.  But the Sun itself tells us nothing of its age, and it is only from having Solar System material that we can analyze in the laboratory that such an exquisite result comes.  For no other star can we do likewise.

Here is a scheme (from \citealt{soderblom:2010}) for thinking about stellar ages, with five quality levels:
\begin{enumerate}
\item {\bf Fundamental} age-dating methods, in which the physics is well understood and all the necessary quantities can be measured.  The only fundamental age is that for the Sun just described.
\item {\bf Semi-fundamental} methods that use well-understood physics, but which require some assumptions because not all the needed quantities are accessible to observation.  These methods include nucleocosmochronology, for very old stars, and kinematic traceback, for young groups.
\item {\bf Model-dependent} methods that, like most of stellar astrophysics, infer ages from detailed stellar models that are calibrated against the Sun.  Model-dependent methods include fitting the loci of stars in clusters (especially the main sequence turn-off) and detecting the lithium depletion boundary (at the transition from stars to brown dwarfs for clusters and groups of stars), as well as isochrone placement and asteroseismology for individual objects.
\item {\bf Empirical} methods that use an observable quantity such as rotation, activity, or lithium abundance that is seen to change with age but for which the physical mechanisms are not understood.  Empirical methods are calibrated against model-dependent ages, generally by observing stars in open clusters.  The paucity of open clusters older than $\sim1/2$ Gyr makes calibrating empirical indicators problematic.
\item {\bf Statistical} methods that rely on broad trends of quantities with age.   
\end{enumerate}

Before going further, I would like to narrow the discussion to methods relevant to Galactic problems as well as methods that produce useful results.  First, in this context we are not interested in very young stars and so kinematic traceback, the lithium depletion boundary, and lithium depletion in general will be ignored.  For a review about ages of young stars, see \cite{soderblom:2014}.  Second, nucleocosmochronology in principle is ideal for age-dating the very oldest stars in the Galaxy.  It uses observations of isotopes of U and Th with long half-lives, but assumptions must be made about the starting abundances, generally done from other r-process elements.  The results often differ by factors of two for the same star, and, in addition, the method needs excellent spectra of very high resolution to detect the weak U/Th features and so has been applied to only a handful of stars \citep{soderblom:2010}.  Third, there have been statistical studies of age-metallicity and age-velocity relations for the Galaxy, but they are at best very noisy relations that are hard to apply to individual objects, and in some cases their reality has been called into question.  They will not be discussed further here.

\section{Ages for Galactic studies}
\label{sec:2}

There are both some special problems and potential advantages when it comes to estimating ages in order to understand the nature of our Galaxy.  These naturally break down into two regimes based on the size scale being considered.  In both cases one wishes to work with large samples of stars that can be selected with completeness or at least for which the biases are known.  I will discuss small-scale samples (the solar neighborhood) as well as Galaxy-wide samples.

\subsection{The solar neighborhood}
\label{subsec:21}

Different studies use various concepts of ``solar neighborhood'' that depend on context.  Here I will consider a radius of $\sim100$ pc.  Within that distance, extinction and reddening are largely negligible; also there are only a very few clusters, and so field stars dominate.  Solar-type stars (F, G, and K dwarfs) make an especially good population for studying this nearby portion of the Galaxy because they are reasonably bright and so identifiable within this sphere, and numerous enough to form a statistically useful sample.  Also, the Sun has a main sequence lifetime of about 10 Gyr, and so G dwarfs are present from all epochs of star formation in the thin disk at least, and in the thick disk as well for late-G stars.  Because they are similar to the Sun and have many narrow lines, G dwarfs yield abundances that are both more precise and more exact than for other stars.  G dwarfs are amenable to age determinations in several ways.  In particular, rotation and activity are known to decline with age, the problem being calibrating the relations, as described below.

The {\it Hipparcos} data make it easy to identify the G dwarfs within 50 pc, for instance, and there are about 3,000 stars from F8V to K2V, enough to make a useful statistical sample.  Because {\it Hipparcos} was brightness limited, there are some biases in its sample: The sample favors binaries, and the completeness horizon diminishes in going to later spectral types, leading to smaller sample sizes.

\subsection{The Galaxy as a whole}
\label{subsec:22}

If we want to delineate the structure of our Galaxy we need to use inherently luminous stars that occur everywhere and which can be seen at the far reaches, even when extincted.   Red giants and stars on the asymptotic giant branch (AGB) fit that description well.  In addition, for Galactic studies it is not necessary to always determine the ages of an entire sample, as one would like to do for, say, exoplanet hosts.  Instead, Galactic work can be done in bulk with a finite failure rate as long as biases are known and can be controlled for.  Finally, Galactic studies ideally need to determine stellar ages on an industrial scale, meaning 10,000 or more.

For these evolved, luminous stars, different age indicators are needed compared to the solar neighborhood sample.  Rotation and activity are useless, as is isochrone placement because all the red giants pile up into a clump.  But, as we will see, asteroseismology is very promising.

At intermediate distances (up to a few kpc), intermediate-mass stars can be helpful.  Stars with main-sequence lifetimes of $\sim$2-4 Gyr (i.e., $\sim$1.5-2$M_\odot$) are promising because many are known and they evolve quickly enough to be placed on isochrones well.

\section{Methods for solar-type stars}
\label{sec:3}

The available methods of age-dating for solar-type stars are discussed in detail in \cite{soderblom:2010} and here I will present only a brief synopsis with recent advances, particularly from asteroseismology.

\subsection{Rotation and activity}
\label{subsec:31}

The trend of declining rotation with age for stars like the Sun has been known for some time \citep{skumanich:1972} and at least has a general scenario to explain it, if not a full  physical model.  In that scenario, it is the fact that the Sun carries energy through convection in its outer layers that leads to essentially all the aspects of the Sun that make it and similar stars ``solar-like," and which make the Sun interesting.  Convection and rotation -- especially differential rotation -- in the Sun's ionized outer layers interact to create a dynamo that regenerates a magnetic field.  This field can grip an ionized stellar wind beyond the stellar surface and thereby transmit angular momentum to it, leading to spindown.  Moreover, more rapid rotation produces a stronger magnetic field and thus more rapid spin down, leading to convergence in the rotation rates of a coeval sample that starts with a spread in initial angular momenta.

This convergence is seen, but it takes at least $\sim500$ Myr (i.e., the age of the Hyades) to occur at $\sim1 M_\odot$ and longer for lower masses.  Until that convergence occurs, each star has its own rotational history and the large scatter seen among young stars in clusters makes it impossible to get an age from rotation.

Past the point of convergence it is assumed that stars obey the $\tau^{-1/2}$ relation ($\tau$ = age) of \cite{skumanich:1972}, but, in fact, we have only the Sun as a well-defined anchor point and so the relation is poorly calibrated.  Adding additional uncertainty, it is possible for a star's companion to add angular momentum late in its life if, say, a close-in giant planet comes close enough to a star to have tidal effects or to even be consumed by the star, adding the orbital angular momentum of the planet to the rotation of the star, causing significant spin-up.  Thus it is possible for some old stars to masquerade as young stars without revealing their true nature.

The situation for activity is worse because activity is an observed manifestation of the magnetic field, which is an indirect consequence of rotation.  In addition, activity varies to some degree on all time scales, from flares and faculae, to rotational modulation by spots, to long-term activity cycles.  As with rotation, the inherent spread in activity among young stars is large.  It is possible to construct a mean activity-age relation \citep{soderblom:1991} that looks good, but averages can hide a lot of inherent noise.

Activity can be fairly easy to measure, requiring spectra with $R\sim 4000$ for the Ca {\sc ii} H and K lines, for instance, making it feasible to get data for many stars quickly.  It is the interpretation of activity that is difficult and uncertain.

X-rays are another manifestation of activity, one with an especially high-contrast signature, but generally x-rays are only detected for the most active stars, which is to say those which are on the ZAMS or younger or which are in close binaries.

In summary, both rotation and activity are flawed as age indicators.  The better of the two is rotation, but rotation versus age remains poorly calibrated for older stars, and there is no calibration at all for metal-poor stars.  In addition, systematic effects can skew rotation rates, and it is not always possible to detect a rotation period for a star, even with very high quality photometry.

\subsection{Isochrone placement}
\label{subsec:32}

Isochrones are loci of constant age computed from stellar models.  Those models are fundamentally calibrated against the Sun, with further refinement from matching details of cluster H-R diagrams.  Fitting isochrones to clusters to determine an age involves using a distribution of stars of many masses, even for sparse clusters or groups.  Even then, fitting the turn-off region can be challenging because there may be only a few stars or the presence of binaries among them may be unappreciated.

One works with less information in placing individual stars in HRDs and so the uncertainties are greater.  As an example, \cite{jorgensen:2005} showed a synthetic study of isochrone placement for F stars.  They took stars of a variety of masses and ages from about 1.2$M_\odot$ on up, added realistic estimates of errors in luminosity and temperature, and then tested how well the original ages could be recovered.  A significant problem in using isochrones is that they are not evenly spaced in the HRD, leading to biased results.  Of the methods they tried, that using Bayesian techniques worked the best because that takes account of prior information about isochrone spacing.  For fairly well-evolved stars above 1.2$M_\odot$, the age errors were as small as 20\%, but were more typically 50\% for most stars.  These are large for individual stars but could be acceptable for ensembles where averaging will reduce the uncertainty.  Another limitation of the Bayesian method is that one ends up with a probability distribution function (PDF), not a specific value.  In many cases the PDF is single-peaked, but there are instances when the PDFs have multiple peaks or give only upper- or lower limits, and, depending on how those are used they can distort or bias an average.

Isochrone placement remains the most favored technique for more massive stars ($>1.5M_\odot$, say), where the relative errors are modest.

\subsection{Asteroseismology}
\label{subsec:33}

For individual stars, asteroseismology offers great promise as a way of determining precise ages, particularly for older stars.  Asteroseismology is essentially like using isochrones, but with significantly better physical constraints.  When asteroseismology works, it works very well indeed, yielding ages as precise as 5\%.  But for the method to work one must detect the oscillations in the first instance, and that requires special assets in almost all cases. Prior to CoRoT and {\it Kepler}, about half a dozen solar-type stars had asteroseismic ages, and each required coordinated observations at several observatories to avoid the problems caused by limited time sampling at just one facility (the diurnal side lobes in the power spectrum).  CoRoT and {\it Kepler} have changed that, but only for stars in their specific fields.

In other words, we do not have a ready means to undertake asteroseismology for any star we'd like.  Additionally, even with {\it Kepler} and its high-quality data one does not always detect the oscillations because the amplitudes are just too low.  Many solar-type stars in the {\it Kepler} field were observed at the one-minute cadence for asteroseismic analysis, but those actually detected tend to be systematically more massive than the Sun and are evolved off the ZAMS.  There are a few detections of oscillations at and below $1M_\odot$, but the number is very small.

Most asteroseismic analysis has been boutique, in the sense that models are calculated for individual stars, one at a time, a labor-intensive process.  Galactic studies in particular will require ages at wholesale ($\sim10^3-10^4$ stars) or even industrial scales ($>10^5$).  An attempt has been made to analyze {\it Kepler} data at the retail ($\sim100$ star) level \citep{chaplin:2013}, where a pipeline and scaling relations are used in place of individual tailoring.  In that case age errors were 25-35\%, with a significant portion coming from inherent differences in different stellar models, even though these stars are all similar to the Sun.

\section{Age-dating evolved stars}
\label{sec:4}

Evolved stars are attractive for Galactic studies because they can be seen at great distances.  They are also favored for determining fairly precise ages.  Red giants exhibit solar-like oscillations that arise from the same mechanism as in the Sun, but the amplitudes are much larger and the periods longer.  For example, the most luminous evolved stars on the asymptotic giant branch (AGB) have amplitudes up to $\sim1$ mmag.

For these stars, detection of the oscillations effectively indicates the mass, and from the mass one can get the age from the known main sequence lifetime.  A large-scale survey at high photometric precision would be needed to detect the oscillations.

\section{Prospects for improvements}
\label{sec:5}

As noted, asteroseismology offers great promise in the area of stellar ages.  When the oscillations can be detected, an analysis using models appropriate to the star can yield ages good to 5-10\% in favorable cases.  This works especially well for older stars, which is ideal since we lack old clusters to calibrate empirical age indicators against.  In particular, stars in the {\it Kepler} field with detected oscillations may be very helpful, even if there are only a few of them.  Oscillations should be easily detectable for evolved stars, at least with space-based data, and those are especially helpful for studying the Galaxy.  The main limitation to seismology is simply being able to obtain data of the necessary quality for the star or stars that one would like.

The detected oscillations of solar-type stars are p (pressure) modes.  Such stars also expected to have g (gravity) modes in their deep interiors.  The g modes penetrate the stellar core, the one part of the star that is a true chronometer.  So far all attempts to detect the Sun's g modes have failed, although some evolved stars in the {\it Kepler} data exhibit mixed modes because there are g-mode frequencies that are at or close to p-mode frequencies.  In those cases the detected modes yield very precise ages.  The ability to consistently detect g modes in stars would enable consistently precise ages.

Gaia won't get parallaxes for every star in our Galaxy, but it certainly should determine extremely accurate distances for all the clusters out to $\sim2$ kpc, and there are many of those with a broad range of inherent properties such as metallicity.  In so doing, the Gaia data essentially remove all distance ambiguity, and that greatly reduces uncertainty about extinction and reddening.  These clusters will enable much more stringent tests of stellar models because of the range of composition available, which should at least range from [Fe/H] = $-1$ to +0.4.  Also, by getting such good distances to clusters, Gaia may indirectly help us understand helium in stars much better than we do now.

Gaia will greatly reduce uncertainty in luminosity for individual stars out to at least $\sim1$ kpc, and that will improve ages from isochrone placement.  However, errors in $T_{\rm eff}$ have been difficult to get below 50 K, and at that level the temperature error dominates the uncertainty in age.  A good way is needed to measure $T_{\rm eff}$ more precisely and reliably.

%\input{referenc}
% BibTeX users please use
%\bibliographystyle{../aa.bst}
%\bibliography{Soderblom_sesto.bib}

% Mac users: please ignore the error message: "! Package natbib Error: Bibliography not compatible with author-year citations."
\end{document}